\documentclass[runningheads,a4paper,final,10pt]{llncs}

\usepackage{tikz}
\usepackage{amsmath}
\usepackage{hyperref}

\usepackage{csquotes}

\usepackage{makecell}

\usepackage{colortbl}

\usepackage{multirow}

\usepackage{booktabs}

\usepackage{subcaption}

\usepackage{caption}

\usepackage{capt-of}

\usepackage{svg}

\usepackage{pifont}
\usepackage{pdflscape}

\usepackage{fp}
\newcommand\printpercent[2]{\FPeval\result{round(#1*100/#2,0)}\result\%}
\newcommand\printpercentfull[2]{\FPeval\result{round(#1*100/#2,2)}\result\%}

\usepackage{ifthen}

\newboolean{anonymous}
\setboolean{anonymous}{True} 
\setboolean{anonymous}{False} 
\newcommand{\anon}[1]{\ifthenelse{\boolean{anonymous}}{#1}{}}
\newcommand{\nonanon}[1]{\ifthenelse{\boolean{anonymous}}{}{#1}}
\newcommand{\ifanon}[2]{\ifthenelse{\boolean{anonymous}}{#1}{#2}}

\newcommand*\numcircledtikz[1]{\tikz[baseline=(char.base)]{\node[shape=circle,draw,inner sep=0.2pt] (char) {#1};}}

\usepackage
{todonotes}
\setuptodonotes{inline}

\usepackage{enumitem}

\newcommand{\reEntry}[1]{\fontsize{8}{9}\selectfont\texttt{#1}}

\newcommand{\numCCActiveCerts}{$1631$}
\newcommand{\numCCArchivedCerts}{$3725$}

\newcommand{\numCcConvertAttempts}{$10\,712$}
\newcommand{\numCcOcrAttempted}{$118$}
\newcommand{\numCcOcrSuccess}{$105$}
\newcommand{\numccschemes}{$17$}
\newcommand{\numccschemeidrules}{$25$}

\newcommand{\evalCcPrecision}{$89\%$}
\newcommand{\evalCcRatioErrorFree}{$81\%$}

\newcommand{\numcccerts}{5356}
\newcommand{\numccids}{5279}
\newcommand{\numccmissingid}{77}

\newcommand{\numccmissingidunfixable}{32}
\newcommand{\numccduplicateid}{151}
\newcommand{\numccduplicateidcolission}{92}

\newcommand{\numccideval}{100}
\newcommand{\numccidevalcorrect}{99}

\newcommand{\numCcRefEval}{100}

\newcommand{\numCCActiveVulnerable}{$88$}
\newcommand{\numCCArchivedVulnerable}{$274$}
\newcommand{\numCCCpeRich}{684}
\newcommand{\fractionCPEShort}{$4.14\%$}  
\newcommand{\numRegularExpressions}{$472$}
\newcommand{\numRegularExpressionGroups}{$33$}
\newcommand{\numSmartcardsVulnerable}{$23$}
\newcommand{\numSmartcardsDistinctVulnerabilities}{$2$}
\newcommand{\numSmartcardsRocaVuln}{$16$}
\newcommand{\numCertsReferencingRocaVuln}{$119$}
\newcommand{\numSmartcardsTitanVuln}{$7$}

\newcommand{\numVulnerableCertsWithMaintenance}{$25$}
\newcommand{\numCertsRevokedWithinYear}{$43$}
\newcommand{\numVulnerableCertsRevokedWithinYear}{$2$}
\newcommand{\numVulnerableCertsBeforeCertification}{$41\%$}
\newcommand{\numVulnerableCertsAfterCertification}{$59\%$}
\newcommand{\numVulnerableCertsInValidityPeriod}{$40\%$}

\begin{document}

\date{}

\title{\Large \bf sec-certs: Examining the security certification practice for better vulnerability mitigation}

\titlerunning{sec-certs}

\nonanon{
\author{Adam Janovsky, Jan Jancar, Petr Svenda, {\L}ukasz Chmielewski, Jiri Michalik, Vashek Matyas}}

\nonanon{
\institute{Masaryk University, Brno, Czechia \email{}}}

\nonanon{
\authorrunning{Janovsky et al.}}

\maketitle

\begin{abstract}
Products certified under security certification frameworks such as Common Criteria undergo significant scrutiny during the costly certification process. Yet, critical vulnerabilities, including private key recovery (ROCA, Minerva, TPM-Fail...), get discovered in certified products with high assurance levels. Furthermore, assessing which certified products are impacted by such vulnerabilities is complicated due to the large amount of unstructured certification-related data and unclear relationships between the certified products.
To address these problems, we conducted a large-scale automated analysis of Common Criteria certificates. We trained unsupervised models to learn which vulnerabilities from NIST's National Vulnerability Database impact existing certified products and how certified products reference each other. Our tooling automates the analysis of tens of thousands of certification-related documents, extracting machine-readable features where manual analysis is unattainable. Further, we identify the security requirements that are associated with products being affected by fewer and less severe vulnerabilities. This indicates which aspects of certification correlate with higher security. We demonstrate how our tool can be used for better vulnerability mitigation on four case studies of known, high-profile vulnerabilities.
All tools and continuously updated results are available at \url{https://seccerts.org} \anon{(web content anonymized)}.

\end{abstract}

\section{Introduction}
Security certification frameworks such as Common Criteria (CC) \cite{cc}, FIPS 140 \cite{fips-140-2}, or EMVCo \cite{emvco} were introduced to independently evaluate the security of products and to provide increased assurance about security-relevant claims for the products' users. The frameworks offer multiple certification levels (e.g., Evaluation Assurance Levels for CC) with increasing requirements and scrutiny undertaken to verify the claims made about the product. Such evaluation comes at a non-trivial cost, both in time to obtain a certificate (months or even years) and in the required finances (hundreds of thousands of dollars or more) \cite{cost_cc,cc_overview_2017}.

Almost immediately after the CC scheme's deployment around the year 2000 (and similarly for the other schemes), discussion about its 
efficacy and limitations emerged and continues until today~\cite{hearn_does_2004,razzazi_common_2006,kaluvuri_quantitative_2014,swcertinpractice_2017}. The main aspects questioned are
\emph{(i)} the inadequate trade-off between the financial or time demands of the certification and the resulting assurance,
\emph{(ii)} difficult orientation of the end users in the ecosystem,
\emph{(iii)} certification being valid only for a fixed product configuration, and 
\emph{(iv)} an unfair position of the vendors with insufficient resources for the certification. Furthermore, critical vulnerabilities \cite{roca,minerva,titan,tpmfail} have been discovered in certified products with high assurance levels, which questions the very purpose of the certification. Worse, assessing their impact on issued certificates and certified products is arduous, due to the large amount of unstructured certification-related data and unclear relationships between certified products.

Existing research on security certifications targets, mostly qualitatively, specific products or a domain \cite{DBLP:conf/ccs/CohneyGH18}, presents anecdotal evidence about the procedural efficiency \cite{swcertinpractice_2017}, or performs a limited quantitative analysis (e.g., only EAL4+ products in \cite{kaluvuri_quantitative_2014}). A systematic insight into the whole certification practice and the security of certified products is missing. Worse, no research work studies the references between the certified products (e.g., composite or updated product) and their effect on vulnerability impact. In this work, we address this gap by automatically assessing all Common Criteria certificates and their artifacts to deliver the following contributions:

\begin{itemize}[noitemsep]
    \item An open-source framework for automated collection and feature extraction of security-relevant information from CC and FIPS 140 certification artifacts. Using our tool, we acquire weekly snapshots of CC and FIPS 140 certificates.
    \item An unsupervised method for mapping vulnerabilities from the National Vulnerability Database (NVD) to affected certified products.
    \item A quantitative analysis of the vulnerabilities impacting CC-certified products, driven by the mapping of  certificates to NVD.
    \item An unsupervised method for learning the inter-certificate references in all CC certificates. The resulting reference graph provides means for vulnerability impact assessment and timely notifications to the stakeholders.
    \item Multiple case studies of how our framework can be leveraged to reveal the true impact of vulnerabilities found in certified products.
\end{itemize}

The paper is organized as follows: Brief introduction to Common Criteria procedures and available artifacts is given in Section~\ref{sec:background} with related work discussed in Section~\ref{sec:related_work}.
Our methodology is described in Section~\ref{sec:methodology}. In Section~\ref{sec:security_impact} we analyze the security impact of various aspects of certification. Section~\ref{sec:helping-vuln-research} presents case studies from real vulnerabilities. We discuss the 
recommendations to improve certification frameworks in Section~\ref{sec:recomm}. Limitations inherent to our work are presented in Section~\ref{sec:limitations}. Finally, we conclude the paper in Section~\ref{sec:conclusions}.

\section{Background}
\label{sec:background}


In the CC certification standard \cite{cc}, any security-related product can be certified with the primary focus 
being the development process. The main assumption is that if the process is correct, the products will most likely be secure. 
Before certification, the applicant needs to choose a set of Security Functional Requirements (SFR) and Security Assurance Requirements (SAR) that the Target of Evaluation (ToE) aims to adhere to. 
Usually, the applicant chooses from an already established list of Protection Profiles (PP) that bind typical use cases (e.g., smartcards) to their usual SFR and SAR specifications. 
During the certification, an independent evaluator, usually an independent laboratory, verifies the product's conformance to this specification. 
Depending on the chosen Protection Profile, 
the product can be certified on multiple Evaluation Assurance Levels (EAL) \cite{cc_overview_2017}, where usually higher EALs provide higher confidence (they also come at a higher cost) that the security features are reliably implemented. 

Within the CC framework, the certification process is performed by many national CC Certificate Authorizing Schemes, and the certification is based on the security evaluation conducted by accredited security laboratories.

For our automated analysis, we use the following documents that accompany the CC certification process and are publicly available:
\begin{description}[noitemsep]
    \item[Certificate] -- issued by the certification scheme, including the most important information, e.g, ToE name and the obtained EAL.
    \item[Certification Report] -- issued by the certification scheme (e.g., US NIAP) after evaluation by an accredited evaluation facility (e.g., Acumen Security), contains a summary of the certification results.
    \item[Security Target] -- provided by the applicant to the evaluation facility, specifies the certified product, including the protection profile, SFR, and SAR. 
    \item[Maintenance Report(s)] -- documenting smaller changes in an already certified product that do not require a full re-certification.
    \item[Protection Profile(s)] -- template for a specific class of products, created either by a single vendor or, more commonly, in collaboration \cite{pp_colab}.
    \item[CSV/HTML web pages]with additional metadata and summary documents automatically generated by the CC portal \cite{cc_webpage} or portals of Certificate Authorizing Schemes. 
\end{description}

We combine the information 
from the aforementioned documents with publicly available information about vulnerabilities. 
In particular, the NVD enriches CVE vulnerability records by, among others, 
connecting vulnerabilities to their vulnerable Common Platform Enumeration (CPE) records \cite{nvd_cpe}. The CPE records aim to fingerprint the product configuration (vendor, product name, version, 
hardware, etc.) and are manually assigned to NVD-listed vulnerabilities. 

\section{Related Work}
\label{sec:related_work}

We divide the studied papers into three different categories: \emph{(i)} papers that identify problems in certification schemes, often suggesting improvements; \emph{(ii)} papers that try to make sense of the certification process, often case-study driven; \emph{(iii)} more recent papers that work towards automated processing of security-related documents, possibly leading to automated and transparent certification.  
Most of the research on CC and FIPS 140 certification schemes builds on individual case studies, which also limits the range of results, as the authors are mostly steered by intuition, not backed with large-scale data analysis. An exception is a study~\cite{kaluvuri_quantitative_2014} that exploits data-driven analysis of CC scheme, parsing metadata from its website. The study, however, focuses only on a subset of certificates, omits the analysis of the certification-related documents, and resorts to manual labor when examining vulnerabilities. Being limited only to EAL4+ certificates forming only around 40\% of the dataset at the time of the their paper writing, the results are relevant only for the domain of ICs and smartcards, which differs significantly from the other categories as demonstrated by our analysis.

In~\cite{hearn_does_2004}, Hearn expresses the challenges that the CC faced around 2004. Eight years later, Murdoch et al.~\cite{murdoch_how_2012} discuss why both CC and FIPS 140 fall short of their promise, pointing to the lack of transparency, a problem that still prevails. More of a political perspective is taken in~\cite{kallberg_common_2012} that describes the fragility of international standards relying on national alliances that may not last forever. Beckert et al.~\cite{Beckert2010MindTG} studies the intriguing fact that CC requires formal verification of the specification to achieve high EALs. Still, the implementation is not formally verified, despite the recent advances in the field. Many of the schemes' problems articulated in our work are also mentioned in a user study~\cite{haney_organizational_2018} of 29 users trained in NIST cryptographic standards and validation programs.

To many, CC and FIPS 140 appear so complex and opaque that it is difficult to understand how one can obtain a certificate and what implications the certification has. This is tackled by several papers that explain the certification process or provide guidance on how to approach it~\cite{cc_overview_2017,mellado_common_2007,herrmann_using_2003,razzazi_common_2006,vetterling_secure_2002,kacprzyk_patterns_2011}. Some works focus on individual product domains for which they provide support (e.g., IoT devices)~\cite{rogowski_2014,bialas_common_2010,kaluvuri_bringing_2013,kang_how_2017}. A study by Shapiro~\cite{shapiro_understanding_2003} dissects the CC-certified (EAL4) Windows version from 2004 and demonstrates the limited practical meaning of EAL levels without considering actual evaluated usage scenario (non-hostile and well-managed user community in this case). A recent paper from 2019 studies the costs introduced by securing software, targeting CC as well~\cite{venson_costing_2019}.


We believe that much precious information is hidden in the PDF artifacts of security certifications. Recent advances in the area of Natural Language Processing (NLP) could be leveraged to translate therein achieved results into the world of information security. This was already suggested for FIPS 140 
by Vassilev~\cite{vassilev2019bowtie}, and performed for the privacy-policies domain by Harkous et al.~\cite{polisis}. Similarly, a study from 2014~\cite{riaz_hidden_2014} attempts to automatically identify security-related sentences and security requirements from natural language artifacts using machine learning. Translating these results into the area of security certificates could allow for automated analysis of potentially vulnerable configurations. In~\cite{DBLP:conf/ccs/CohneyGH18}, Cohney et al. demonstrated that manual analysis of FIPS 140 security policy documents can yield discoveries of critically flawed PRNG configurations.

In our work, we rely on the mapping between CVEs and the associated CPEs as provided by the NVD. But even constructing such mapping solicits active research. The respective papers try to make this process more efficient and complete, or less erroneous~\cite{tovarnak_graph_2021,wareus_automated_2020,sanguino_software_2017}.

A closed-source tool called \emph{CCScraper}~\cite{ccscraper-2022} 
developed by the jtsec security lab since 2018 automatically analyzes information from the CC and Certification Scheme portals using OCR capabilities and other features. While the tool seems similar to our \texttt{sec-certs}, there are notable differences -- it is not oriented at certification process evaluation using the NVD database, and it does not seem to produce a graph of references usable for vulnerability notification. As the tool is not publicly available, we cannot provide a more detailed comparison. 

In contrast to prior work, we fully automate the analysis of certification artifacts. Better, we learn the references between the certificates and map the certified products to existing vulnerabilities, quantitatively exploring the association between the level of certification requirements and severity of vulnerabilities found on all items certified under Common Criteria. 

\section{Methodology} \label{sec:methodology}

\begin{figure*}[ht]
    \captionsetup{font=footnotesize}
    \centering
    \includegraphics[width=1.0\textwidth]{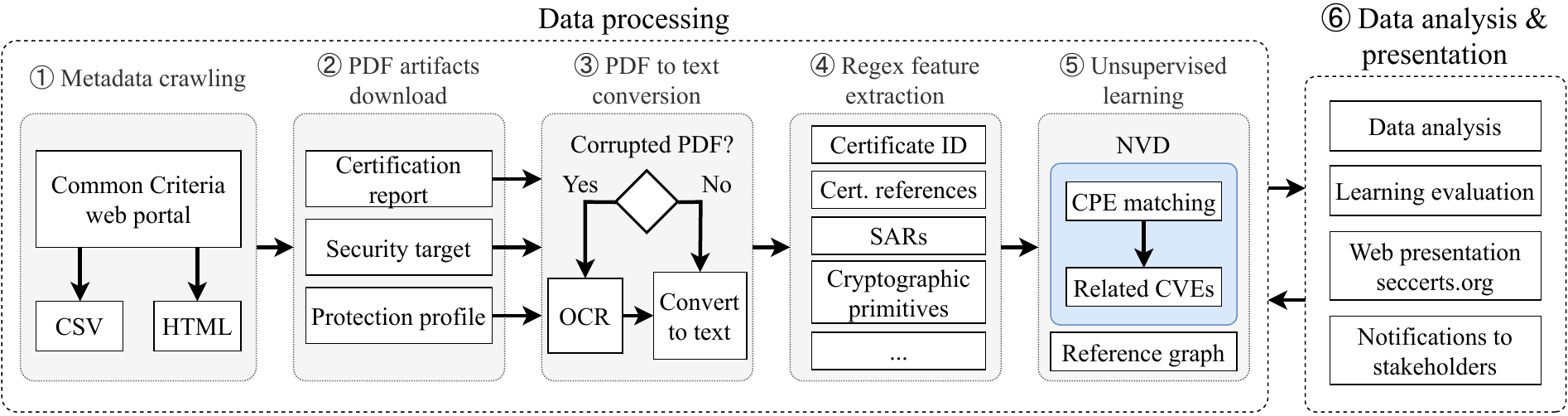}
    \caption{A high-level overview of the \texttt{sec-certs} analysis framework.}
    \label{fig:methodology:framework}
\end{figure*}

In this section, we describe the methodology of our work, starting with a high-level description of our framework that is depicted in Figure~\ref{fig:methodology:framework}. To explore the landscape of CC-certified products, we collect, dissect and analyze the certification artifacts in the form of varyingly structured PDF documents and accompanying metadata. We begin by crawling the CC website (\numcircledtikz{1}), from which we extract links to certification artifacts along with various metadata. Next, we download (\numcircledtikz{2}) the PDF documents and convert them to text (\numcircledtikz{3}). When converted, we extract features from the text using regular expressions (\numcircledtikz{4}). Together with extracted metadata, we feed the extracted attributes into unsupervised learning models (\numcircledtikz{5}) to learn references between certificates and to link the certified products with the NVD. Finally, we evaluate these models, analyze the collected statistics and present the results to stakeholders on our web (\numcircledtikz{6}).

We release our open-source framework
\footnote{\ifanon{Code: \url{https://anonymous.4open.science/r/sec-certs-7A92}, web: \url{https://seccerts.org} (web content anonymized)}{Code: \url{https://github.com/crocs-muni/sec-certs}, web: \url{https://seccerts.org}}.}
that implements all of these components. It is written in Python and also contains code for analysis of FIPS 140-2/3 certificates. On our web, we publish weekly fresh snapshots of our dataset to capture newly issued certificates and monitor possible changes in the artifacts. Our web offers an easy way to interact with our results and we envision it as an aid in vulnerability impact assessment as described later in Section~\ref{sec:helping-vuln-research}.

In the rest of this section, we examine the individual layers in detail.

\subsection{Metadata crawling \& PDF download} \label{subsec:metadata_crawling}

Every national scheme must regularly and publicly disclose a list of certified products as mandated by the Arrangement on the Recognition of Common Criteria Certificates in the Field of IT Security~\cite{ccra_2014}. Those lists are consolidated and made available on the Common Criteria Portal (\url{commoncriteriaportal.org}), which serves as the authoritative source for certification documents. The portal provides information on certified products in two formats: a CSV list and detailed HTML pages. We automatically aggregate data from both the CSV and HTML formats, creating a unified entry for each certified product. To this end, we crawled records of \numCCActiveCerts{} active and \numCCArchivedCerts{} archived certificates. Despite a handful of inconsistencies between the CSV and HTML data, we consider the obtained records to be the ground truth as they were obtained from the official CC portal. For each certificate, we obtain URLs of the following PDF documents that we download and further analyze: \emph{(i)} certification report, \emph{(ii)} security target, and \emph{(iii)} maintenance update(s).

\subsection{PDF conversion \& feature extraction} \label{sec:methodology:subsec:pdf_to_text_conversion}

The PDF artifacts vary in their form between vendors and schemes. To enable fast processing of the documents, we convert them from PDF to text using the \texttt{pdftotext} library~\cite{pdftotext}. An alternative library \texttt{PyMuPDF}~\cite{pymupdf} was considered as a candidate solution. The conversion results for 100 random certification artifacts were qualitatively compared by a researcher with industrial NLP experience. Based on this experiment, \texttt{pdftotext} was chosen with \texttt{--raw} flag that preserves the text in content stream order. Not all documents convert correctly; issues in conversion lead us to examine certain characteristics, such as file size or character ratios. To this end, we detect \numCcOcrAttempted{} documents out of \numCcConvertAttempts{} as problematic (the complete list of performed checks is presented in Appendix~\ref{app:conversion_options}). We correct the conversion of the malformed documents with the \texttt{Tesseract} OCR library~\cite{smith2007overview}, succeeding in \numCcOcrSuccess{} ($97\%$) cases. 

After converting the artifacts to text, we extract keywords from them using \numRegularExpressions{} regular expressions in \numRegularExpressionGroups{} groups. We devised these expressions using expert knowledge to cover broad areas, ranging from certificate identifiers through cryptographic algorithms to standards. A list of the categories, along with example matches, is available in Table~\ref{tab:regex-categories}. The complete list of expressions is accessible from our repository\footnote{\ifanon{\url{https://anonymous.4open.science/r/sec-certs-7A92/src/sec_certs/rules.yaml}}{\url{https://github.com/crocs-muni/sec-certs/blob/main/src/sec_certs/rules.yaml}}}. We evaluate the quality of the involved regular expression by monitoring the output quality of the corresponding algorithms.

\begin{table}[ht]
\centering
\begin{tabular}{lr@{\hskip 7pt}l@{\hskip 7pt}r@{\hskip 7pt}lr}
    \fontsize{8}{9}\selectfont\# & Category & Example & Category & Example & \fontsize{8}{9}\selectfont\# \\\toprule
    & \multicolumn{4}{c}{\textbf{Common Criteria}} & \\\midrule
    \fontsize{8}{9}\selectfont \numccschemeidrules{} & Certificate ID & \reEntry{ANSSI-CC-2012/35} & Evaluation level & \reEntry{EAL2} & \fontsize{8}{9}\selectfont 3 \\
    \fontsize{8}{9}\selectfont 8 & Protection Profile ID & \reEntry{BSI-CC-PP-0072-2012} & SAR & \reEntry{AVA\_VLA.4} & \fontsize{8}{9}\selectfont 12 \\
    \fontsize{8}{9}\selectfont 61 & Evaluation facility & \reEntry{Riscure} & SFR & \reEntry{FCS\_COP.1} & \fontsize{8}{9}\selectfont 11 \\
    \fontsize{8}{9}\selectfont 5 & Certification process & \reEntry{out-of-scope} & Claim & \reEntry{T.DOC\_OPEN} & \fontsize{8}{9}\selectfont 10 \\\midrule
    & \multicolumn{4}{c}{\textbf{Cryptography}} & \\\midrule
    \fontsize{8}{9}\selectfont 56 & Symmetric crypto. & \reEntry{AES} & Hash-function & \reEntry{SHA-256} & \fontsize{8}{9}\selectfont 27 \\
    \fontsize{8}{9}\selectfont 21 & Post-quantum crypto. & \reEntry{KYBER} & Scheme & \reEntry{PKE} & \fontsize{8}{9}\selectfont 7 \\
    \fontsize{8}{9}\selectfont 10 & Asymmetric crypto. & \reEntry{RSA-2048} & Protocol & \reEntry{TLS} & \fontsize{8}{9}\selectfont 9 \\
    \fontsize{8}{9}\selectfont 6 & Randomness source & \reEntry{RNG} & Cipher mode & \reEntry{CBC} & \fontsize{8}{9}\selectfont 12 \\
    \fontsize{8}{9}\selectfont 17 & Elliptic curve & \reEntry{P-256} & Library & \reEntry{OpenSSL} & \fontsize{8}{9}\selectfont 26 \\
    \fontsize{8}{9}\selectfont 1 & TLS cipher suite & \reEntry{TLS\_AES\_256\_GCM\_SHA384} & Engine & \reEntry{SmartMX2} & \fontsize{8}{9}\selectfont 3 \\\midrule
    & \multicolumn{4}{c}{\textbf{Device}} & \\\midrule
    \fontsize{8}{9}\selectfont 20 & Vendor & \reEntry{NXP} & Device model & \reEntry{STM32F446ZCT} & \fontsize{8}{9}\selectfont 4 \\
    \fontsize{8}{9}\selectfont 19 & Trusted exec. env. & \reEntry{ARM TrustZone} & OS & \reEntry{JCOP 4} & \fontsize{8}{9}\selectfont 2 \\
    \fontsize{8}{9}\selectfont 7 & JavaCard package & \reEntry{java.security.*} & IC data group & \reEntry{EF.DG15} & \fontsize{8}{9}\selectfont 5 \\
    \fontsize{8}{9}\selectfont 32 & JavaCard constant & \reEntry{ALG\_AES\_CBC\_PKCS5} & CPLC data & \reEntry{IC.Fabricator} & \fontsize{8}{9}\selectfont 3 \\
    \fontsize{8}{9}\selectfont 5 & JavaCard version & \reEntry{JavaCard 2.2} & & \\\midrule
    & \multicolumn{4}{c}{\textbf{Miscellaneous}} & \\\midrule
    \fontsize{8}{9}\selectfont 24 & Side-channel attack & \reEntry{DPA} & Vulnerability & \reEntry{CVE-2017-15361} & \fontsize{8}{9}\selectfont 5 \\
    \fontsize{8}{9}\selectfont 14 & Standard & \reEntry{NIST SP 800-90A} & Tech. report & \reEntry{BSI TR-03110} & \fontsize{8}{9}\selectfont 2 \\
    \bottomrule
\end{tabular}
\vspace*{0.2cm}
\caption{Regular expression categories - as used in our feature extraction - and their example matches, the numbers denote the number of regular expressions in each category, in total there are \numRegularExpressions{}.}
\label{tab:regex-categories}
\end{table}

\subsection{Unsupervised learning} \label{sec:methodology:subsec:unsupervised_learning}

We use the features extracted by the previous stages of the pipeline to tackle two different problems: \emph{(i)} mapping of certified products to known vulnerabilities affectin them, and \emph{(ii)} building a graph of references between certified products. We cast these as classification problems and we propose two unsupervised algorithms to tackle them. To evaluate the quality of our models, we manually visited at least 100 annotations produced by each model. The co-authors assessed these to measure the models' precision -- the proportion of true positives relative to the sum of true positives and false positives. This metric is as a critical indicator of our models' accuracy in identifying correct solutions, thereby ensuring the reliability of our approach in handling both problems.

\subsubsection{Mapping certified products to vulnerabilities.}

We designed a classifier that maps each certified product to a (possibly empty) list of vulnerabilities from the National Vulnerability Database that likely affect it. A prior work~\cite{kaluvuri_quantitative_2014}, manually mapped a small fraction of certificates to their CVEs. That approach, however, does not scale and cannot be automatically applied to newly discovered vulnerabilities. In contrast, our method is fully automated and works with previously unseen certificates and vulnerabilities.

\begin{figure*}[t!]
    \captionsetup{font=footnotesize}
    \centering
    \includegraphics[width=1.0\textwidth]{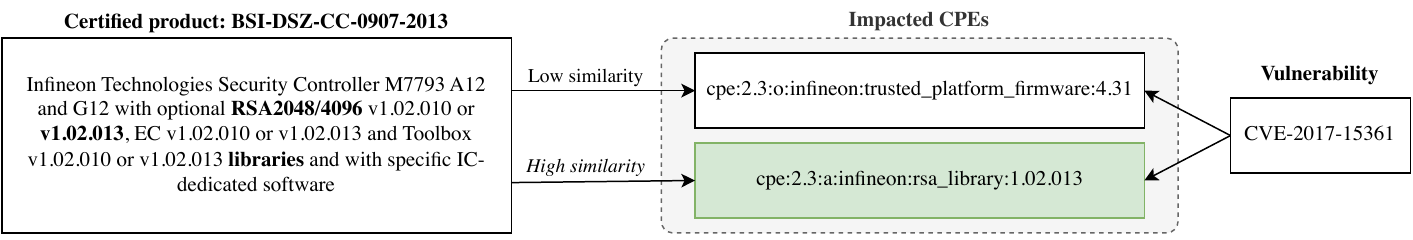}
    \caption{An example of mapping a certified product to a vulnerability that impacts it. While \texttt{<vendor>} fields in both CPEs match the vendor of the certified product, only the  {\color{green!40!black}highlighted CPE} has its \texttt{<version>} field appearing in the certified product title. Moreover, the string similarity between the lemmatized certificate title (turning the token \enquote{librarires} into its base form \enquote{library}) yields a perfect match. Thus, the certified product matches the CPE and is likely impacted by the vulnerability.}
    \label{fig:cpe_matching_example}
\end{figure*}

Every CVE in the NVD is associated with a manually crafted specification of vulnerable product configurations listed as Common Platform Enumeration (CPE) records. Given a CPE record of a product, one can scan the NVD to find vulnerabilities that affect it. To learn what vulnerabilities affect certified products, it thus suffices to create a mapping between CPEs and Common Criteria certificates, a process illustrated in Figure~\ref{fig:cpe_matching_example}. We consider the mapping between CPEs and CVEs maintained by NIST as being error-free, i.e., that each CVE record contains exactly the CPE records that correspond to the vulnerable configurations.

CPE is a data structure consisting of fields describing a product configuration. Among others, every CPE contains the fields \texttt{<vendor>}, \texttt{<product>}, and \texttt{<version>}~\cite{waltermire_technical_2018}. To match the certified products to CPEs, we measure the string similarity between a \textit{lemmatized} certificate's title and candidate CPE records\footnote{We attempted to enhance our approach by matching certificate names directly to the vulnerability descriptions, but this fell short of its promise, delivering deteriorated performance.}. We empirically fine-tuned the required similarity threshold to achieve nearly 90\% precision of the mapping on the expert-annotated samples. While achieving higher precision is possible, it asymmetrically decreases the recall, i.e., the number of matched CPE records.

To measure how similar a CPE record is to a given certificate's title, we use the fuzzy metrics from RapidFuzz package~\cite{max_bachmann_2021_5584996}. The used metric is the \emph{indel distance}, the minimal number of character insertions/deletions that convert one string to the other. Multiple variants of indel distance exist, each suitable for different scenarios. We use two different variants of this metric and return their maximum:

\begin{enumerate}
    \item \emph{Partial token sort ratio}: sort the words in each string alphabetically, find the best alignment of the shorter string in the longer string and return the normalized indel distance for this alignment.
    \item \emph{Partial token set ratio}: Find all alphanumeric tokens in each string and treat them as sets. Rebuild the two strings into the common form \texttt{<sorted intersection><sorted remainder>}. Compute the normalized indel distance for this pair. 
\end{enumerate}

We require three conditions to hold before considering any CPE to be a candidate for a match: \emph{(i)} The CPE \texttt{<product>} field must be at least four characters long\footnote{Short CPEs lead to 100\% string similarity, this step discards \fractionCPEShort{} CPE records.} ; \emph{(ii)} vendor fields of the certificate and the CPE must match perfectly; \emph{(iii)} the major and minor versions extracted from the certificate title must match those in the CPE. For example, matching a certified product of version \texttt{5.1.2} to a CPE record with version \texttt{5.1} is allowed. This reflects the intuition that a vulnerability affecting a platform of version \texttt{5.1} also likely affects version \texttt{5.1.2}. 

For implementation details, we refer the reader to our repository. The naive representation of our algorithm that maps a single certificate to its corresponding CPEs is as follows:

\begin{enumerate}
    \item Use regular expressions to extract product versions; leverage data from CC portal to recover the manufacturer and the full title of the certified product.
    \item Select candidate CPEs from the set of all CPEs such that:
    \begin{itemize}
        \item[(i)] Their \texttt{<product>} field is at least 4 characters long.
        \item[(ii)] Their \texttt{<vendor>} field is identical to the manufacturer of the certified product.
        \item[(iii)] Their major and minor version match the versions extracted from the certified product.
    \end{itemize}
    \item Measure the string similarity between the CPE and the product's title using the discussed fuzzy metrics. If above threshold, declare the CPE as a match, and the certified product as likely affected by the vulnerabilities impacting the CPE. 
\end{enumerate}

\subsubsection{Building a graph of references.}
In the CC framework, every certificate (except a handful of exceptions) is assigned its own supposedly unique but structurally heterogeneous identifier (ID) under the national issuing scheme (e.g., \textit{BSI-DSZ-CC-1169-2021} or \textit{ANSSI-CC-2017/24}). Notably, the CC framework does \emph{not} provide an authoritative pairing of certified products and IDs in public documents, nor does it provide a list of all certificate IDs, nor does it prescribe any form of these IDs. The only common characteristic of IDs across national schemes is the presence of \emph{some} counter. Often, the IDs contain the year of certification, sometimes they even identify the evaluation facility or the vendor. For a list of example IDs, see Table~\ref{tab:certid-examples}.

Certificates often reference each other in their artifacts. The reference signals some component relationship among the certified products, such as a smartcard referencing a certificate of a cryptographic chip used within it. Such references are vital for assessing the reach of a newly found vulnerability in a certified product. For example, a vulnerability in a certified cryptographic chip may affect other certified products, such as eID cards, as witnessed during the ROCA vulnerability case~\cite{roca}. To aid this vulnerability assessment, we retrieve the full graph of inter-certificate references from the artifacts. First, we recover a unique ID of each certificate and then we search all other certificates for the presence of such ID. The ID is assigned as follows:

\begin{enumerate}[noitemsep]
    \item Extract all candidate certificate IDs using \numccschemeidrules{} regular expressions for the \numccschemes{} schemes from the: \emph{(i)} filename, \emph{(ii)} PDF metadata, \emph{(iii)} frontpage and \emph{(iv)} contents, of the certification report of a given certificate. Assign a weight to each candidate equal to its number of occurrences, normalized by source.
    \item Filter candidates to match the scheme of the certificate.
    \item Canonicalize candidates, to correct errors and join variants of the same ID.
    \item Merge canonical candidates from the four sources into one list, weighing the sources by \texttt{1.0} \emph{(i)}, \texttt{1.2} \emph{(ii)}, \texttt{1.5} \emph{(iii)} and \texttt{1.0} \emph{(iv)} respectively, and summing the weights of duplicate candidates.
    \item Pick the candidate with the largest weight as the ID of the certificate. If there are more, pick the longest.
\end{enumerate}

We constructed the regular expressions for matching the certificate IDs by an iterative process where we repeatedly examined certification reports that we did not yet have a matching ID for and broadened our regular expression set based on the certificate IDs we manually identified. This led to the list of \numccschemeidrules{} regular expressions, which are also available in Appendix \ref{app:certid_regexes}. To canonicalize the certificate IDs, we use the same regular expressions first to parse individual ID components (such as the year or counter) and then construct the canonical ID with a manually developed scheme-specific method. The rest of the heuristics applied, such as the weighting and preferences for longest candidate IDs, were devised after testing and experimentation to correct erroneous assignments that we observed.

\begin{table}[ht]
    \centering
    \vspace*{-0.2cm}
    \begin{tabular}{cl|cl}
        \toprule
        \textbf{Scheme} & \textbf{ID} & \textbf{Scheme} & \textbf{ID}\\\midrule
         AU & \texttt{Certificate Number: {\color{red}2008}/{\color{blue}49}} & CA & \texttt{{\color{blue}522} {\color{orange!80!black}EWA} {\color{red}2020}} \\
         DE & \texttt{BSI-DSZ-CC-{\color{blue}1052}-{\color{green!40!black}V3}-{\color{red}2021}} & ES & \texttt{{\color{red}2018}-6-INF-{\color{blue}2529}-{\color{green!40!black}v2}} \\
         FR & \texttt{ANSSI-CC-{\color{red}2018}/{\color{blue}57}{\color{green!40!black}v2}} & IN & \texttt{IC3S/{\color{orange!80!black}KOL01}/ADVA/EAL2/{\color{blue}0520}/{\color{blue}0022}} \\
         IT & \texttt{OCSI/CERT/{\color{orange!80!black}TEC}/{\color{blue}02}/{\color{red}2009}/RC} & JP & \texttt{JISEC-CC-CRP-C{\color{blue}0599}-01-{\color{red}2018}} \\
         KR & \texttt{KECS-NISS-{\color{blue}0612}-{\color{red}2015}} & MY & \texttt{ISCB-5-RPT-C{\color{blue}104}-CR-{\color{green!40!black}V1a}} \\
         NL & \texttt{NSCIB-CC-{\color{red}17}-{\color{blue}67206}-CR{\color{green!40!black}2}} & NO & \texttt{SERTIT-{\color{blue}040}} \\
         SE & \texttt{CSEC{\color{red}2016}{\color{blue}012}} & SG & \texttt{CSA\_CC\_{\color{red}21}{\color{blue}005}} \\
         TR & \texttt{21.0.03/TSE-CCCS-{\color{blue}41}} & UK & \texttt{CRP{\color{blue}225}} \\
         US & \texttt{CCEVS-VR-{\color{red}03}-{\color{blue}0044}} & & \\\bottomrule
    \end{tabular}
    \vspace*{0.2cm}
    \caption{Example certificate IDs of Common Criteria schemes. Note that certain schemes may have multiple formats, with only one example presented here. Legend: {\color{blue}counter}, {\color{red}year}, {\color{green!40!black}version}, {\color{orange!80!black}evaluation facility}. }
    \label{tab:certid-examples}
    \vspace*{-0.5cm}
\end{table}

\subsubsection{Model evaluation.}
\label{subsec:eval_refs_graph}

\begin{table}[ht]
    \centering
    \begin{tabular}{l|c|c|c}
    \toprule
    \textbf{Problem} & \textbf{Precision} & {\textbf{Evaluation set size}} & {\textbf{\fontsize{8}{9}\selectfont\# annotated certs.}} \\
    \midrule
    CPE matching & \evalCcPrecision{} & 100 & \numCCCpeRich{} (\printpercent{\numCCCpeRich{}}{\numcccerts{}}) \\
    Certificate ID matching & \printpercent{\numccidevalcorrect{}}{\numccideval{}} & \numccideval{} & \numccids{} (\printpercent{\numccids{}}{\numcccerts{}}) \\
    \bottomrule
    \end{tabular}
    \vspace*{0.2cm}
    \caption{Evaluating our unsupervised models: The size of the evaluation set indicates how many instances received expert annotations for precision calculation.  The final column depicts the number of certificates that our model positively annotated. It is important to note that not all products have corresponding CPEs.}
    \label{tab:model_performance}
\end{table}

We manually verified the models' outputs on a sampled set of certificates to evaluate the quality of our algorithms. Table~\ref{tab:model_performance} shows our models' comparison.

For the problem of CPE matching, we randomly sampled 100 certificates for which the model predicted at least one CPE. Two co-authors then independently annotated each predicted match as true positive or false positive. The conflicting annotations were revisited by the pair of co-authors until consensus was reached.
This enabled us to measure the precision of the classifier. That is especially relevant for CPE matching, where our results form a conservative lower bound, with probably more CPEs (and therefore vulnerabilities) applicable to the certified products. When our model declared a CPE match, it was correct in \evalCcPrecision{} cases on this expert-annotated subset -- furthermore, \evalCcRatioErrorFree{} certificates contained completely error-free matches.

For the problem of reference graph building, we manually annotated a random set of \numccideval{} certificates with their certificate IDs. This was done by one expert co-author (as there is little space for ambiguity) using the certification artifacts. We then compared the manual annotations with those assigned by our model. We observed that our classifier achieved \printpercent{\numccidevalcorrect{}}{\numccideval{}} precision. Of the cases where our model did not assign an ID (\numccmissingid{}, \printpercentfull{\numccmissingid{}}{\numcccerts{}} of the dataset) \numccmissingidunfixable{} of the errors are not fixable, mostly due to the certification artifacts containing no recognizable ID or due to the certification artifacts not being available. We also investigated certificates where our method assigned the same ID to multiple certificates, of which there were \numccduplicateid{}. Upon further analysis, \numccduplicateidcolission{} of these were due to multiple certificates sharing the same certification report files (i.e., duplicates in the input data from the CC portal).

\subsection{Data analysis \& presentation} \label{subsec:data-analysis}

We run this processing pipeline weekly to capture newly certified products and observe possible changes in the artifacts. We release all processed data on \url{seccerts.org} to offer an easy way to interact with and browse through our results. We envision the portal as an aid in vulnerability impact assessment as we describe later in Section~\ref{sec:helping-vuln-research}. Our web delivers a plethora of functionalities:

\textbf{Certificates:} Users can browse through certificates, and see the extracted data in a concise and visually rich way. Local copies of the certification artifacts in PDF, as well as textual forms, are also accessible. The certificate page also renders an interactive graph of references.

\textbf{Search and filtering:} The site provides both title-based and full-text search of the certification artifacts. Such search functionality is a powerful tool for impact assessment of vulnerabilities, as will be demonstrated in Section~\ref{sec:helping-vuln-research}.

\textbf{Data:} The machine-readable data extracted using our methodology can be downloaded from the site. Our Python framework also integrates this download mechanism so users can use up-to-date processed data for their custom analysis without needing to run the data-extraction pipeline.

\textbf{Notifications:} Users can subscribe to get email updates
when the data associated with a certificate changes. This way, a user can be notified of new vulnerabilities that may potentially affect certificates of their interest.


\section{Measuring security impact of certification} \label{sec:security_impact}

In this section, we apply the classifier described in Section~\ref{sec:methodology} that maps certificates to vulnerabilities to examine relations between vulnerabilities and the certified products. With our analysis, we aim to answer several intriguing questions: When, in the life-cycle of the certified product do the vulnerabilities tend to appear? What are the most common weaknesses that affect certified products? Does the distribution of the weaknesses differ between open-source projects and certified products? What is the role of the maintenance updates (a formal step in certification process) in vulnerability remediation? And finally, is number and severity of the vulnerabilities associated with the degree of scrutiny that a product undergoes during the certification process? 

Before we answer these questions, we first assess our data resource, the National Vulnerability Database. NVD is arguably the most widely used database of vulnerabilities. Better, it is manually curated and vetted. As such, NVD has proven a central resource for assessing vulnerability impact and remediation in various ecosystems~\cite{security-patches,security-vulns-python}, and its data backs many security tools\footnote{Such security tools include: Lacework, Snyk, Veracode, or others.}. Still, it must be understood that a small fraction of records may exhibit inconsistencies~\cite{cleaning-nvd} and that not all exploitable vulnerabilities may get their CVE issued. During our investigation, we noticed that this is the case, especially for the smartcard industry that does not consider the CVE database as the primary means for vulnerability reporting\footnote{Search for various forms of \enquote{smartcard} in NVD returns only handful of results.}. In our dataset, we register a major discrepancy between the number of certified smartcards and a number of certified smartcards with a vulnerability. In fact, we detected only \numSmartcardsDistinctVulnerabilities{} distinct vulnerabilities for this category with \numSmartcardsVulnerable{} vulnerable certificates in total. \numSmartcardsRocaVuln{} certified smart cards suffer from the ROCA~\cite{roca} and \numSmartcardsTitanVuln{} from Titan~\cite{titan_cve} vulnerability, both high-profile flaws. For that reason, smartcards are further omitted from the rest of this section. We affirm that the ratio of CVE-rich certified products to all certified products is uniform in \emph{all other categories}.

\begin{table}[b]
\scalebox{0.7}{
\begin{minipage}[b]{0.7\linewidth}
\centering
\begin{tabular}{ c | c | c }
\toprule
\textbf{CWE-ID} & \textbf{CWE name} & \textbf{\# CVEs} \\
\midrule
CWE-119	 & Buffer overflow 	 & 892 \\
CWE-20	 & Improper Input Validation 	 & 487 \\
CWE-200	 & Sensitive information exposure 	 & 349 \\
CWE-264	 & Access control error 	 & 316 \\
CWE-787	 & Out-of-bounds Write 	 & 297 \\
CWE-125	 & Out-of-bounds Read 	 & 208 \\
CWE-399	 & Resource Management Errors 	 & 180 \\
CWE-79	 & Cross-site Scripting 	 & 148 \\
CWE-416	 & Use After Free 	 & 122 \\
CWE-362	 & Race Condition 	 & 115 \\
\bottomrule
\end{tabular}
\vspace{2mm}
\caption{Top 10 weaknesses (CWEs) identified from 3114 vulnerabilities that affect 362 distinct CC-certified products.}
\label{tab:cwes}
\end{minipage}\hfill
\hspace{0.2cm}
\begin{minipage}[b]{0.68\linewidth}

\input{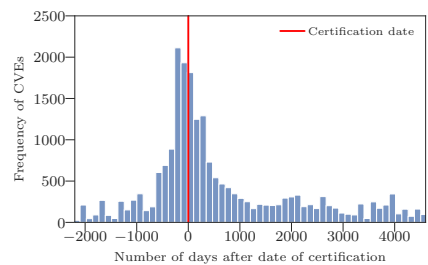}
\captionof{figure}{Disclosure dates of CVEs that affect certified products in relation to the certification date.}
\label{fig:vuln_histogram}
\end{minipage}
}
\end{table}

\subsection{Vulnerabilities in certified products}

Among \numCCActiveCerts{} active (\numCCArchivedCerts{} archived) certificates, we identified \numCCActiveVulnerable{} (\numCCArchivedVulnerable) with at least one assigned vulnerability. As can be seen from Figure~\ref{fig:vuln_histogram}, approximately \numVulnerableCertsBeforeCertification{} of vulnerabilities were publicly disclosed in the NVD \emph{before} the certification date. This means that the majority (\numVulnerableCertsAfterCertification{}) of vulnerabilities were disclosed after the security evaluation related to CC certification process. Worse, if a certified product got affected by a CVE, in \numVulnerableCertsInValidityPeriod{} cases it was during its certificate's lifetime. In the rest of this section, we study only the vulnerabilities disclosed during the validity period of the product.

We also investigated certificates with a validity period shorter than 365 days. While it is not transparent whether a certificate was withdrawn or set to expire that soon (this information cannot be publicly backtracked), it is highly unlikely that a certificate would be intended to be valid for less than a year. In our data, we identified \numCertsRevokedWithinYear{} such certificates. Yet, we register a CVE only for \numVulnerableCertsRevokedWithinYear{} of them. We, therefore, conjecture that the vendors revoke the certificates early for reasons mostly unrelated to \emph{publicly} disclosed vulnerabilities.

\subsection{Most common weaknesses}

Many of the vulnerabilities in the NVD are manually assigned a set of so-called common weaknesses from the Common Weakness Enumeration (CWE)~\cite{cwe} that capture the cause and nature of the vulnerability. We linked the vulnerabilities occurring in CC certificates to their weaknesses, learning what weaknesses cause most harm in the landscape of certified products
. Table~\ref{tab:cwes} shows a diverse set of weaknesses affecting certified products with buffer overflow, improper input validation, and sensitive information exposure being the most common. The prevalence of weaknesses in our dataset does not change dramatically over time and the distribution is comparable to the one reported in the landscape of open-source projects by Bhandari et al.~\cite{bhandari2021cvefixes} (all their top-8 CWEs are present in our top-15 list). This experiment shows that the certified products suffer from weaknesses similar to those identified in the open-source projects.

\subsection{Higher EAL, fewer vulnerabilities?}

The CC standard categorizes~\cite{cc_sars_2017} Security Assurance Requirements into multiple classes that convey different aspects of product security. For instance, the \texttt{ATE} class stands for \textit{tests}, whereas the \texttt{AVA} class stands for \textit{vulnerability assessment}. The classes are further split into families. For example, \texttt{ATE\_COV} specifies test coverage while \texttt{ATE\_DPT} specifies test depth. Each of these families has a different domain of levels with free-text descriptions. For example, \texttt{ATE\_COV} can be of levels 1, 2, or 3, depending on the test coverage during the development. These fine-grained requirements are the basis for more coarse EALs.

\begin{table*}[t]
\captionsetup{font=scriptsize}
\centering
{\scriptsize
\begin{tabular}{ c | c c | c c | c c }
    \toprule
    & \multicolumn{2}{ c |}{\emph{number of CVEs}} &  \multicolumn{2}{ c |}{\emph{Average CVE base score}} & \multicolumn{2}{ c }{\emph{variable characteristics}} \\
    variable & $\rho$ & p-value & $\rho$ & p-value & support & domain range \\
    \midrule
EAL &  -0.02 &  1.64e-01 & \cellcolor{green!15} -0.28 & \cellcolor{green!15} 4.24e-05 & 2927 & 14 \\
ALC\_CMS & \cellcolor{green!15} -0.07 & \cellcolor{green!15} 7.34e-05 & \cellcolor{green!15} -0.35 & \cellcolor{green!15} 2.55e-07 & 2583 & 5 \\
ALC\_CMC & \cellcolor{green!15} -0.08 & \cellcolor{green!15} 3.54e-05 & \cellcolor{green!15} -0.35 & \cellcolor{green!15} 3.05e-07 & 2578 & 5 \\
AVA\_VAN & \cellcolor{green!15} -0.08 & \cellcolor{green!15} 1.29e-05 & \cellcolor{green!15} -0.34 & \cellcolor{green!15} 6.47e-07 & 2606 & 4 \\
ATE\_COV &  -0.03 &  8.50e-02 & \cellcolor{green!15} -0.33 & \cellcolor{green!15} 8.41e-06 & 2419 & 3 \\
ATE\_IND & \cellcolor{green!15} -0.05 & \cellcolor{green!15} 1.51e-03 & \cellcolor{green!15} -0.29 & \cellcolor{green!15} 1.54e-06 & 3321 & 3 \\
ADV\_FSP & \cellcolor{green!15} -0.06 & \cellcolor{green!15} 4.19e-04 & \cellcolor{green!15} -0.24 & \cellcolor{green!15} 6.00e-05 & 3371 & 6 \\
ASE\_OBJ & \cellcolor{green!15} -0.07 & \cellcolor{green!15} 1.33e-03 &  -0.14 &  4.46e-02 & 2080 & 2 \\
ASE\_REQ & \cellcolor{green!15} -0.07 & \cellcolor{green!15} 2.94e-04 &  -0.12 &  6.01e-02 & 2120 & 2 \\
ALC\_FLR &  0.06 &  9.94e-01 &  0.14 &  9.52e-01 & 1684 & 3 \\
    \bottomrule
\end{tabular}
}
\vspace*{0.5em}
\caption{Spearman's rank correlations between SARs (EAL) and two attributes extracted from the certified product: number of CVEs and the average CVE base score. The \emph{support} column denotes the sample size for the given pair of variables. The \emph{domain range} 
displays the number of distinct ranks empirically measured in the given SAR or EAL. The p-value is associated with the null hypothesis \emph{\enquote{The examined variables are not negatively correlated}}. Associations with p-value $<0.01$ (higher SAR level likely leads to less severe vulnerabilities) are highlighted. 
We report 
a full list of the examined associations during our study.}
\label{tab:correlations}
\vspace*{-1em}
\end{table*}

It is natural to hypothesize that products validated to higher assurance levels will suffer from fewer and less severe vulnerabilities. In other words, the levels of SARs (EAL) should be negatively associated with those variables. Using our tool, we reconstructed the SAR levels of a certificate from three sources: \emph{(i)} assurance level field in the CC-provided CSV file, \emph{(ii)} keywords in the Security Target, \emph{(iii)} keywords in the Certification Report. Based on this information, we established a set of certificates' SARs. We then measured the association between SARs (EAL) and two variables: \emph{(i)} Number of CVEs that the product suffered from \textit{during its validity period}; \emph{(ii)} Average \textit{base score} (severity) of the related vulnerabilities (from the Common Vulnerability Scoring System~\cite{cvss}). Given the ordinal nature of the variables, Spearman's rank correlation coefficient ($\rho$) was used to measure the association. To have sufficient sample sizes, we focused on SARs with $>100$ vulnerable certificates and a diverse domain\footnote{At least two SAR levels found, each in $>40$ certificates.}, which yielded an analysis of $9$ distinct SARs. All reported findings were computed on the confidence level $0.99$ using a one-sided alternative hypothesis that the variables \emph{are negatively} correlated.  

We report the following results: 7/9 SARs are negatively associated with the number of CVEs affecting the given certificate. The associations, however, are very weak, with $\left | \rho \right | \leq 0.07$. More importantly, 6/9 SARs are negatively associated with the average base score of the affecting vulnerabilities, showing a strong relationship with $0.24 \leq \left | \rho \right | \leq 0.35$. These are: functional specification, independent testing, test coverage, vulnerability assessment, configuration management capabilities, and configuration management scope. Interestingly, EAL is correlated only with the average base score of the vulnerabilities, with $\rho = - 0.28$. We did not encounter a positive association during our experiments. This experiment shows what security requirements have the highest association with lower average CVE base scores, possibly preventing catastrophic bugs. 



\subsection{Do maintenance update fix CVEs?}



In CC, the vendors can participate in so-called assurance continuity~\cite{assurance_continuity} that enables making minor or major changes to the ToE without repeating the whole product evaluation. The bug fixes to ToE can be released in so-called \emph{maintenance updates} if they are categorized as minor changes. The the analysis of the bug is, however, hidden in undisclosed Impact Analysis Report (IAR). 

We searched for all certificates that exhibited at least one vulnerability \textit{after they were certified and before a maintenance update was released}, manually investigating the contents of the respective update documents to learn their rationale. This yielded \numVulnerableCertsWithMaintenance{} updates to study that all conveyed some vulnerability analysis. Three encountered scenarios were particularly troubling. First, one update contains the phrase \enquote{The guidance document has been updated to ensure a vulnerable feature, which is disabled by default, remains disabled.}. This suggests that the vulnerable feature remains in the product, and the user is instructed not to trigger it. Second, one of the products used a maintenance update (issued 3 years after certification) to fix 3 CVEs that were publicly disclosed before the product was \emph{initially} certified, showing a considerable delay. And third, multiple updates argued that the detected vulnerabilities fall out of ToE's scope, with the rationale explained in the IAR that is not available for public scrutiny. 


 \section{Improving vulnerability mitigation} \label{sec:helping-vuln-research}
Our methodology and data improve the impact assessment of vulnerabilities discovered in certified products. There are several benefits to certification ecosystem stakeholders: end-users can get early notification of a potential vulnerability for the products they use, security researchers can shortlist candidates for further scrutiny and responsible disclosure, and vendors can assess their certified products. Importantly, end-users are not dependent solely on the information flow from the vulnerable product vendor, which may have misaligned incentives hindering timely notification or may not be able to directly contact the end-user -- as is the case when a product is sold via intermediaries.

To demonstrate the contributions of our tool, we analyze four past vulnerabilities and compare the information obtained using \texttt{sec-certs} with a mature understanding of the vulnerabilities. We cover  different types of vulnerabilities: algorithmic vulnerability in RSA keypair generation in Infineon cryptographic chips (ROCA \cite{roca}), timing-leakage vulnerabilities in STM chips and several cryptographic libraries (TPMFail \cite{tpmfail} and Minerva \cite{minerva}), and an EM-leakage vulnerability in several smartcards (A Side Journey to Titan \cite{titan}).

We note that such an evaluation is qualitative rather than quantitative. It is undoubtedly easier to search for something knowing it exists, i.e., affected products. The evaluation might be biased due to the vulnerability type (we are investigating only several high-profile vulnerabilities) or the affected product type (we focus on the domain of cryptographic hardware where we expect higher benefits due to overall ecosystem secrecy and obscurity).

\subsection{ROCA (CVE-2017-15361)} \label{subsec:case-roca}

To demonstrate our method, we begin with the case of Estonian eID documents utilizing a chip vulnerable to the ROCA attack \cite{roca} with certificate ID \href{https://seccerts.org/cc/350581534e265186/}{ANSSI-CC-2013/55} \cite{estIDCert}. After the responsible disclosure of the vulnerability to Infineon (the manufacturer of the vulnerable chip) by the original researchers, large affected parties were supposed to be notified before the public disclosure, either by the eID vendor or via non-public memos distributed among the parties involved in the EU-wide eIDAS directive. However, the Estonian government received the vulnerability information only at the beginning of September 2017, close to the municipal elections with e-voting utilizing the eID cards with the vulnerable chips. While we cannot assess the information distributed by the vendor directly, we can analyze the information distributed in the eIDAS incident report memo ID 163484 
issued to all EU members on 20th of June 2017 \cite{eidasmemo} by an affected Austrian e-health provider. That memo was insufficient for the Estonian government to become aware of the impact. 

The memo states that \enquote{CardOS V5.0 with Application for QES, V1.0} and \enquote{CardOS V5.3 QES, V1.0} are vulnerable (directly corresponding to certificates \href{https://seccerts.org/cc/5ee9898421c06859/}{BSI-DSZ-CC-0833-2013} and \href{https://seccerts.org/cc/7e8763f553e43d12/}{BSI-DSZ-CC-0921-2014}). Also, a \enquote{weakness in asymmetric crypto library} was mentioned and that the \enquote{problem affects generating electronic signature creation data for use with the RSA algorithm} was noted. A concrete vendor was named in the memo, which was not the manufacturer of the vulnerable chip, but a platform integrator instead. Using these two certificates, we search for references from these certificates, which gives us two chip certificates: \href{https://seccerts.org/cc/8eb6fca41668f95b/}{BSI-DSZ-CC-0758-2012} and \href{https://seccerts.org/cc/9be76c10474e0c80/}{BSI-DSZ-CC-0782-2012}. 
Both certificates are a version of Infineon Security Controller M7892 with \enquote{optional RSA2048/4096 v1.02.013 ... libraries}, matching the weakness description from the memo. Having the vulnerable chip certificates, we can search for certificates  referencing them. More than one hundred certificates are found, yet Estonia's 
\href{https://seccerts.org/cc/350581534e265186/}{ANSSI-CC-2013/55} is not among them as it uses a different certified chip (but with the same cryptographic library). Additionally, the cluster of devices related to these named in the Austrian report is only very indirectly connected to the cluster \href{https://seccerts.org/cc/350581534e265186/}{ANSSI-CC-2013/55} belongs to. To match \href{https://seccerts.org/cc/350581534e265186/}{ANSSI-CC-2013/55}, we need to create the initial set of vulnerable certificates from all chips with the vulnerable \texttt{RSALib v1.02.013} library and build a reference graph from this set -- crucial insight that requires understanding what was shared between the two certified products in the memo, as well as the description of the weakness. Automatic search alone would not be sufficient in this case, but relatively simple refinement with input from a security expert is sufficient.

After the public disclosure of the ROCA vulnerability \cite{roca}, many end-users attempted to determine whether they were impacted. While, in this case, it was possible to detect the vulnerable product directly from the statistical properties of generated public keys, such an option is not readily available in general. The corresponding CVE-2017-15361\footnote{\url{https://nvd.nist.gov/vuln/detail/CVE-2017-15361}} description illustrates the case of incomplete or even misleading information about a specific vulnerability in NVD. While the CVE initially listed the exact version of the vulnerable library, it mentioned only TPM chips with CPEs mostly consisting of affected laptop platforms -- omitting the fact that large majority of Infineon smartcards was also vulnerable as the same library was used.

As the vulnerable library version is publicly known from the CVE description, we 
search for the strings \enquote{RSALib} and \enquote{1.02.013} using the CC portal \cite{cc_webpage}; this yields only 3 and 32 certificates, respectively. In contrast, \texttt{sec-certs} directly links \numSmartcardsRocaVuln{} certificates to the CVE. 
Additionally, we automatically find \numCertsReferencingRocaVuln{} certificates that reference the vulnerable ones, a significant improvement. 


\subsection{TPM-Fail (CVE-2019-16863)} \label{subsec:tpmfail}
The TPM-Fail paper presented a timing attack vulnerability 
\footnote{\url{https://nvd.nist.gov/vuln/detail/CVE-2019-16863}}
on ECDSA signing in several TPM implementations, including a CC and FIPS 140-2 certified chip, the \textit{ST33TPHF2ESPI} with firmware version 73.04 \cite{tpmfail}. The paper points to the \href{https://seccerts.org/cc/10b17081dd7cad8f/}{ANSSI-CC-2018/41} certificate as vulnerable. Also, it links to the product brief of the chip, which advertises CC and FIPS 140-2 certifications, but does not point to concrete certificates. Additionally, the CVE listing from NVD provides several CPEs that include the chip name in few variations and firmware versions.

Using \texttt{sec-certs}, we can search for the chip name and firmware versions, yielding 5 CC certificates, including the one mentioned in the paper. By looking at references of the found certificates, we uncover 7 additional certificates concerning the same family of \textit{ST33TPHF2*} chips. Comparing their reported firmware versions with the CPEs from NVD leads to 6 certificates matching the vulnerable versions, while the original paper reported only the one.

\subsection{Minerva (CVE-2019-15809)}
The Minerva group of timing attack vulnerabilities in ECDSA signing \cite{minerva} coincided with and exploited the same type of leakage as TPM-Fail, but targeted a different set of implementations, including smartcard certified under both CC and FIPS 140-2, the \textit{Athena IDProtect}. Its authors used information from CC and FIPS 140-2 certificates to report on the affected certified products, using manual analysis of the certificate reports and security target documents.

The authors report a single vulnerable CC certificate for the \textit{Athena IDProtect} smartcard with ID \href{https://seccerts.org/cc/fca98ecd003e1b82/}{ANSSI-CC-2012/23}. A concrete affected certificate could be identified as its certification report included the CPLC (Card Production Life Cycle) data that matched the data on the vulnerable smartcard. The authors also report on the root cause of the vulnerability as being in another certified item, the \textit{Atmel Cryptographic Toolbox 00.03.11.05} with certificate ID \href{https://seccerts.org/cc/49b4531177e9c3af/}{DCSSI-2009/11}. The vulnerable card's certificate references this certificate. Its security target describes two sets of functions for performing elliptic curve cryptography operations, \textit{secure} and \textit{fast}, with the fast functions not offering any SPA/DPA protection. As the authors note, it is likely that the vulnerable smartcard mistakenly decided to use the fast functions. This sort of root cause analysis would not be possible without using the certification artifacts of CC certificates. 

Using \texttt{sec-certs}, we can search for known vulnerable firmware and its variations using \textit{00.03.11.0\textbf{*}} wildcard, finding another, previously unreported and likely also impacted version \textit{00.03.11.08}. Additionally, the reference graph reveals 32 potentially vulnerable certificates that directly reference those with vulnerable firmware. 



\subsection{Google Titan  (CVE-2021-3011)}
The Side Journey to Titan \cite{titan} paper demonstrated a side-channel attack on ECDSA signing in a chip used in the \textit{Google Titan} security key, as well as in other secure elements and smartcards. While the \textit{Google Titan} security key used the \textit{NXP A7005a} chip, the authors note that directly attacking a commercial secure element with countermeasures in a black-box way is usually really hard. They thus chose to try to get access to a product with the same underlying cryptographic implementation yet with more control:
\textquote{\textit{We went through the public data that can be found online and figured out that several NXP JavaCard smartcards are based on P5x chips and have similar characteristics with the NXP A700X \cite{titan}.}} The authors then performed a similar analysis of CC certification reports as in the case of Minerva, noting:
\blockquote{\textit{Thanks to BSI and NLNCSA CC public certification reports, we were able to gather the following (non-exhaustive) list of NXP JavaCard smartcards based on P5x chips...}}.

The associated CVE-2021-3011\footnote{\url{https://nvd.nist.gov/vuln/detail/CVE-2021-3011}}
has a listing of vulnerable smartcards. Our \texttt{sec-certs} tool automatically maps the CVE to \numSmartcardsTitanVuln{} CC certificates. Mapping it to the rest of the certificates mentioned in the paper is then a question of looking at the neighborhood of the certificates in the reference graph and perhaps a full-text search query. A re-analysis done using \texttt{sec-certs} found 15 additional certificates referencing the vulnerable cryptographic library (\href{https://seccerts.org/cc/0f684159ad31f883/}{BSI-DSZ-CC-0633-2010}), mostly ePassport products that might be vulnerable.
According to the authors of \cite{titan}, once they validated the attack on NXP JCOP J3D081 and on the Titan key, they spent between half a day to one day manually searching for all the CC certificates of similar products (mainly using the Common Criteria and the BSI website). 
In comparison, our search using the \texttt{sec-certs} tool took 15 minutes. \\

In this section, we revisited four high-profile vulnerabilities and demonstrated how using \texttt{sec-certs} both reduces the time required to perform the vulnerability assessment and leads to a higher number of identified certificates that are affected by the vulnerability. In all cases, incorporating the \texttt{sec-certs} to the researcher's workflow would positively augment the originally achieved results; this comparison is further highlighted in Table~\ref{tab:manual_vs_seccerts}.

\begin{table*}[t]
\captionsetup{font=scriptsize}
\centering
{\scriptsize
\begin{tabular}{ l | c | c | c | c }
    \toprule
    & \multicolumn{2}{c}{\emph{ \# certs. identified }}  &  \multicolumn{2}{c}{\emph{ estimated workload }} \\
    Vulnerability & orig & sec-certs & orig & sec-certs \\
    \midrule
ROCA (CVE-2017-15361) & 32* & 16+119 & not reported & 20 minutes \\
TPM-Fail (CVE-2019-16863) & 1 & 5+7 & not reported & 10 minutes \\
Minerva (CVE-2019-15809) & 2 & 2+32 & 1-2 days & 15 minutes \\
Titan (CVE-2021-3011) & 11 & 7+15 & 0.5-1 day & 15 minutes \\
    \bottomrule
\end{tabular}
}
\vspace*{0.5em}
\caption{Comparison of the workload and the number of identified vulnerable certificates as reported by the first vulnerability reporters in contrast with the analysis done with the \texttt{sec-certs} tool. Note that in all cases, \texttt{sec-certs} was clearly superior both in the number of identified certificates and in the time required to perform the given task. *Number of certificates from ROCA vulnerability was not reported by the authors; full-text search for vulnerable library version '1.02.013' yields 32 certificates.}
\label{tab:manual_vs_seccerts}
\vspace*{-1em}
\end{table*}

\section{Recommendations}
\label{sec:recomm}

Based on the lessons learned through  our paper, we give recommendations that would improve the existing certification frameworks. Better, most of our recommendations require little effort to implement and, when mandated on newly certified products, would arguably improve certification transparency, vulnerability assessment, or automatic processing.

\textbf{Assign robust certificate ID.} Unique and robust IDs assigned to each certificate allow for clear and precise referencing of related certificates. The ID shall also not be too long to make referencing less error-prone. BSI or NSCIB formats are compact and sufficient. A format used by FIPS 140 consisting only of a counter is particularly challenging to extract unambiguously.

\textbf{Specify evaluated product boundaries more clearly.} The Target of Evaluation (ToE) is specified in a somewhat informal way, making it difficult for automatic processing to establish the ToE boundaries. Their definition is now dispersed through the certification documents, with their form prohibiting clear understanding from the stakeholders. The software equivalent of the Bill of Material (BoM) was proposed as a partial solution, but the task is more complicated: ToE covers not only the parts of the certified product but also the circumstances and environment of use. Fuzzy boundaries make it difficult for end users to verify whether the product used is in the certified configuration or to properly pair and evaluate the vulnerabilities reported. Some vendors already supply an automated tool checking if a product is running in a certified configuration. Further, while the evaluation labs act as trusted mediators to evaluate certification claims, the steps taken, tools used, and results obtained by evaluation labs are confidential and not independently replicable later. At least part of the tests shall be automated and made with product results openly available. 

\textbf{Pre-assign platform records (e.g., CPE) to certificates.} Well-defined records remove ambiguity in certificate naming and enable more accurate assignment of vulnerabilities to the affected platform(s), resulting in a complete bidirectional linking between certified products and 
vulnerabilities. Some products, like operating systems, consist of multiple software packages or other components. As the certification procedure already requires the vendor to monitor a vulnerability database for relevant vulnerabilities, the vendor likely has a list of CPEs to watch that could be made public.

\textbf{Define and use pre-defined machine-readable templates}. As the certified products from the same category and/or implementing the same Protection Profile frequently follow a very similar structure, pre-defined templates would not only speed up document preparation but, more importantly, also allow for automated checks against predefined criteria and options selected. Unsurprisingly, such templates are already in use, as visible from the documents prepared by some of the larger vendors, but are not publicly available and cross-compatible.


\textbf{Be more transparent in vulnerability handling.} While the certified products can include patches that protect them from existing vulnerabilities that otherwise affect the product of such a version, the information about whether these patches are applied is not provided. Maintenance updates often neglect vulnerabilities or present generic statements like \emph{\enquote{IAR contains a rationale why these vulnerabilities do not affect the ToE}}. We recommend that IARs get disclosed as a part of the update and that all vulnerabilities considered during the update are explicitly listed with their IDs. Additionally, the IAR should clearly comment on the rationale for considering some vulnerabilities as out-of-scope.

\textbf{Provide a clear reason for certified product validity end.} A reason for a product archival helps end-users to assess its further usage, especially for a product with non-standard or a prematurely ending validity period (e.g., due to unfixable vulnerability or a mistake in certificate issuance). The certification scheme acts here as an independent and trusted party, since the product vendor might be in a conflict of interest to fully and timely disseminate the reason for certificate validity termination. The inspiration can be taken from handling and revoking web PKI certificates, possibly including automated validity check services like Online Certificate Status Protocol (OCSP).  


\textbf{Design certification tests for replicability.} Evaluation labs function as trusted mediators that allow the assessment of certification claims based on proprietary (non-public) documents. But steps taken, tools used, and results obtained by evaluation labs are frequently confidential. Moreover, the level of expertise found in specialized evaluation laboratories is not available to all end users. Public documents, like maintenance reports, also frequently lack the data used during their issuance. As a result, certification and subsequent annual re-evaluations of a product are usually done by the same lab. Due to confidentiality, these activities are not independently replicable later. Test automation and detailed description of the steps taken, preferably with openly available tools, would allow for better replicability and retrospective assessment.

\textbf{Improve data quality served by official certification portals.} The primary sources of certification artifacts \cite{cc_webpage,fips_webpage} contains inaccuracies like improperly formatted CSV files with duplicate rows, non-existent links, missing or incorrect valid until date, duplicated documents, or corrupted files. 

\textbf{Make information available in standardized, machine-readable formats.} Cost of creation of an additional metadata file is likely negligible with respect to overall certification cost but significantly improves the usefulness of certification artifacts after the certification is finished.

\section{Limitations}
\label{sec:limitations}

Our analysis has several limitations, mostly due to incomplete or inconsistent input data. We discuss their possible impact on the presented results below.

\textbf{Noisy data.} The certification documents are a free-form, often ambiguous, and inconsistent data source. We mitigate some errors by augmenting inputs (e.g., CC product lists from both CSV and HTML inputs). Where possible, we measure the precision of our heuristics on manually annotated certificate subsets.


\textbf{Unclear ToE boundaries.} A security target specifies the boundaries of the evaluated product, which typically do not span all product components. Yet, our mapping of certificates to vulnerabilities cannot account for a limited ToE and considers all parts of a product to be inside the ToE. While technically an error, we believe that this generally matches the expectations of end-users. 
The unclear boundary is even more pressing given anecdotal evidence that vendors may purposely limit the ToE scope in order to limit the attack surface during the evaluation instead of improving the security of the product and fixing potential vulnerabilities.
As an example of such issue, consider the certificate \href{https://seccerts.org/cc/16ba6dab2c5c4b13/}{SERTIT-115}~\cite{sertit-115} of a \emph{network} camera. Its network interface was considered to be always trusted and therefore likely untested during its evaluation.\footnote{The certification report~\cite{sertit-115} states: \enquote{Attackers have no chance to connect any malicious devices into the local network of the TOE.}} 

\textbf{Limitations of NVD.} Not all vulnerabilities are assigned a CVE record and/or adequately equipped with affected platform CPEs. Thus, our work forms a lower bound on the number of vulnerabilities in certified products with more vulnerabilities likely existing. 

\textbf{Context-free references.} While many certificates reference others to signal a sub-component relationship, this may not always be the case. As our reference analysis does not distinguish the reference context, false positives are possible when using the graph of references for assessment of vulnerability impact. To measure the scope of this issue, we performed a manual analysis of references in \numCcRefEval{} certificates that we reported in Subsection~\ref{subsec:eval_refs_graph}.

\section{Conclusions and future work}
\label{sec:conclusions}

The evaluation labs responsible for the security certification of products are in a conflict of interest as they are paid by vendors. This incentivises them to certify products that should not otherwise pass the certification process. This issue has already been recognized by Cohney et. \cite{DBLP:conf/ccs/CohneyGH18} who noted that it was one of the reasons for NIST moving away from the \enquote{independent test lab} model. Together with non-transparent certification often based on proprietary documents and procedures, this undermines the credibility of the whole certification ecosystem. 

To assist the stakeholders of Common Criteria, we turned the existing certification artifacts into a machine-processable dataset, paired the certificates with NVD, and reconstructed the full graph of inter-certificate references. By doing so, we gained an insight into various attributes of the CC certificates that span over 25 years. This increases transparency and accountability of the whole certification. Better, we allow the stakeholders to fast and precise impact assessment of vulnerabilities that may affect products of their interest.

We demonstrated that higher rigor in certification requirements -- as conveyed by Security Assurance Requirements (SAR) -- is associated with a \emph{lower} number and \emph{less severe} vulnerabilities for the majority of examined SAR classes. Further, to display the effectiveness of our tool in vulnerability assessment, we revisited four major incidents that struck CC-certified products. The \texttt{sec-certs} tool proved invaluable during such analysis, leading to the identification of many additional vulnerable products, and considerably reducing the time required for the analysis. 

Finally, we deliver a web presentation with weekly updated results of our work, providing the stakeholders with easy access to analyzed certification artifacts and to notifications about newly discovered vulnerabilities that may impact them. This will help to ensure that stakeholders are informed of the latest developments and can make informed decisions about the products they use. The whole toolchain can also be self-hosted and enriched with additional -- potential non-public information -- for stronger security analysis. 

As future work, we envision natural language processing as a valuable tool to extract security-relevant information from the certification artifacts. Cohney et al.~\cite{DBLP:conf/ccs/CohneyGH18} already showed that one can identify vulnerable products only by manually searching for alerting phrases in their text specification, and we believe that such task could be automated and outsourced to recent NLP models. Additionally, NLP models could automatically identify the context of the references between the certified products, or even identify non-certified dependencies (e.g., included cryptographic library). With such ability at hand, it would be possible to more closely analyze the threats to the ecosystem of certified products. Further, we consider expanding our analysis also to EMVCo~\cite{emvco} artifacts.

\section*{Acknowledgments}\label{sec:ack}

This work was supported and funded by the following institutions and grants.  The Internal grant agency of MUNI, CZ.02.2.69/0.0/0.0/19\_073/0016943, RedHat Czech, Cyber Security Network of Competence Centres for Europe - CyberSec4Europe, Tools for AI-enhanced Security Verification of Cryptographic Devices (AI-SecTools VJ02010010), Cyber-security Excellence Hub in Estonia and South Moravia (CHESS, 101087529).





\bibliographystyle{plain}
\bibliography{references.bib}

\appendix

\section{Malformed PDF conversion detection}
\label{app:conversion_options}

\begin{table}[ht]
    \centering
    \begin{tabular}{l|l|c}
    \toprule
    \textbf{Index} & \textbf{Attribute} & \textbf{Threshold}  \\
    \midrule
    1 & Number of lines & $30$ \\
    2 & File size & $1000\textrm{B}$ \\
    3 & Average line length & $20$ \\
    4 & \# lines with even characters non-identical & $15$ \\
    5 & Alphanumeric characters ratio & $0.5$ \\
    \bottomrule
    \end{tabular}
    \vspace*{0.5cm}
    \caption{Various checks are conducted on text files converted from PDF artifacts. If any evaluated attribute falls \emph{below} a set threshold, the conversion is considered flawed, prompting an attempt to correct it using OCR. Check 4 specifically looks for lines where every other character is the same, a common issue in converted ANSSI documents with improper spacing, typically involving repeated spaces. The individual threshold values were devised through empirical testing.}
    \label{tab:garbage_detection}
\end{table}

\begin{landscape}

\vspace*{-2cm}
\section{Certificate ID regular expressions}
\label{app:certid_regexes}

\begin{table}[!h]
    \centering
    \footnotesize
    \vspace*{-0.6cm}
    \hspace*{-2.5cm}
    \begin{tabular}{c|l}
        \toprule
        Scheme & Regex \\
        \midrule
    AU &
\begin{minipage}{8.1in}
\begin{verbatim}
(Certificate Number:|Certification Report) (?P<year>[0-9]{2,4})/(?P<counter>[0-9]+)
\end{verbatim} 
\end{minipage} \\[-0.1em]\cmidrule(r){1-1}\addlinespace[-0.1em]
    \multirow{2}{*}{CA} & 
\begin{minipage}{8.1in}
\begin{verbatim}
(?P<number1>383)[ -](?P<digit>[0-9])[ -](?P<number2>[0-9]+)(-CR|P)?
\end{verbatim} 
\end{minipage} \\
        &
\begin{minipage}{8.1in}
\begin{verbatim}
(?P<number>[0-9]+)[ -](?P<lab>EWA|LSS|CCS)([ -](?P<year>[0-9]+))?
\end{verbatim} 
\end{minipage} \\[-0.1em]\cmidrule(r){1-1}\addlinespace[-0.1em]
        DE & 
\begin{minipage}{8.1in}
\begin{verbatim}
BSI-DSZ-CC-((?P<s>S)-)?(?P<counter>[0-9]{3,5})-?((?P<version>[vV][0-9])-)?(?P<year>[0-9]{4})?(-(?P<doc>(RA|MA)(-[0-9]+)?))?
\end{verbatim}
\end{minipage} \\[-0.1em]\cmidrule(r){1-1}\addlinespace[-0.1em]
    ES &
\begin{minipage}{8.1in}
\begin{verbatim}
(?P<year>[0-9]{4})[-‐](?P<project>[0-9]+)[-‐]INF[-‐](?P<counter>[0-9]+)[ -‐]{1,2}[vV](?P<version>[0-9])
\end{verbatim} 
\end{minipage} \\[-0.1em]\cmidrule(r){1-1}\addlinespace[-0.1em]
        \multirow{4}{*}{FR} & 
\begin{minipage}{8.1in}
\begin{verbatim}
DCSS[Ii]-(?P<year>[0-9]{2,4})/(?P<counter>[0-9]+)([vV](?P<version>[0-9]))?
\end{verbatim} 
\end{minipage} \\
        &
\begin{minipage}{8.1in}
\begin{verbatim}
Rapport de certification (?P<year>[0-9]{2,4})/(?P<counter>[0-9]+)([vV](?P<version>[0-9]))?
\end{verbatim} 
\end{minipage} \\
        &
\begin{minipage}{8.1in}
\begin{verbatim}
Certification Report (?P<year>[0-9]{2,4})/(?P<counter>[0-9]+)([vV](?P<version>[0-9]))?
\end{verbatim} 
\end{minipage} \\
        &
\begin{minipage}{8.1in}
\begin{verbatim}
ANSS[Ii](-CC)?[ -](?P<year>[0-9]{2,4})[/_-](?P<counter>[0-9]+)(-(?P<doc>([MSR][0-9]+)))?([vV](?P<version>[0-9]))?
\end{verbatim} 
\end{minipage} \\[-0.1em]\cmidrule(r){1-1}\addlinespace[-0.1em]
    IN &
\begin{minipage}{8.1in}
\begin{verbatim}
IC3S/(?P<lab>[A-Z]+[0-9]+)/(?P<vendor>[a-zA-Z_]+)/(?P<level>[a-zA-Z0-9]+)/(?P<number1>[0-9]+)/(?P<number2>[0-9]+) ?(/CR)?
\end{verbatim} 
\end{minipage} \\[-0.1em]\cmidrule(r){1-1}\addlinespace[-0.1em]
    IT &
\begin{minipage}{8.1in}
\begin{verbatim}
OCSI/CERT/((?P<lab>[A-Z]{3})/)?(?P<counter>[0-9]{2,3})/(?P<year>[0-9]{4})/RC
\end{verbatim} 
\end{minipage} \\[-0.1em]\cmidrule(r){1-1}\addlinespace[-0.1em]
    \multirow{3}{*}{JP} & 
\begin{minipage}{8.1in}
\begin{verbatim}
(CRP|ACR)-C(?P<counter>[0-9]+)-(?P<digit>[0-9]+)
\end{verbatim} 
\end{minipage} \\
        &
\begin{minipage}{8.1in}
\begin{verbatim}
JISEC-CC-CRP-C(?P<counter>[0-9]+)-(?P<digit>[0-9]+)-(?P<year>[0-9]{4})
\end{verbatim} 
\end{minipage} \\
        &
\begin{minipage}{8.1in}
\begin{verbatim}
Certification No. [cC](?P<counter>[0-9]+)
\end{verbatim} 
\end{minipage} \\[-0.1em]\cmidrule(r){1-1}\addlinespace[-0.1em]
    KR &
\begin{minipage}{8.1in}
\begin{verbatim}
KECS[-‐](?P<word>ISIS|NISS|CISS)[-‐](?P<counter>[0-9]{2,4})[-‐](?P<year>[0-9]{4})
\end{verbatim} 
\end{minipage} \\[-0.1em]\cmidrule(r){1-1}\addlinespace[-0.1em]
    MY &
\begin{minipage}{8.1in}
\begin{verbatim}
ISCB-(?P<digit>[0-9])-RPT-C(?P<counter>[0-9]{3})-CR(-[0-9])?-(?P<version>[vV][0-9][a-z]?)
\end{verbatim} 
\end{minipage} \\[-0.1em]\cmidrule(r){1-1}\addlinespace[-0.1em]
    NL & 
\begin{minipage}{8.1in}
\begin{verbatim}
(NSCIB-|CC-|NSCIB-CC-)(?P<core>((?P<year>[0-9]{2})-)?(-?[0-9]+)+)(-?(?P<doc>(CR|MA|MR)[0-9]*))?
\end{verbatim} 
\end{minipage} \\[-0.1em]\cmidrule(r){1-1}\addlinespace[-0.1em]
    NO & 
\begin{minipage}{8.1in}
\begin{verbatim}
SERTIT-(?P<counter>[0-9]+)
\end{verbatim} 
\end{minipage} \\[-0.1em]\cmidrule(r){1-1}\addlinespace[-0.1em]
    SE &
\begin{minipage}{8.1in}
\begin{verbatim}
CSEC ?(?P<year>[0-9]{4})(?P<counter>[0-9]{2,3})
\end{verbatim} 
\end{minipage} \\[-0.1em]\cmidrule(r){1-1}\addlinespace[-0.1em]
    SG &
\begin{minipage}{8.1in}
\begin{verbatim}
CSA_CC_(?P<year>[0-9]{2})(?P<counter>[0-9]{3})
\end{verbatim} 
\end{minipage} \\[-0.1em]\cmidrule(r){1-1}\addlinespace[-0.1em]
    TR &
\begin{minipage}{8.1in}
\begin{verbatim}
(?P<prefix>[0-9\\.]+)/TSE-CCCS-(?P<number>[0-9]+)
\end{verbatim} 
\end{minipage} \\[-0.1em]\cmidrule(r){1-1}\addlinespace[-0.1em]
        \multirow{2}{*}{UK} & 
\begin{minipage}{8.1in}
\begin{verbatim}
CRP(?P<counter>[0-9]+[A-Z]?)
\end{verbatim} 
\end{minipage} \\
        &
\begin{minipage}{8.1in}
\begin{verbatim}
CERTIFICATION REPORT No. P(?P<counter>[0-9]+[A-Z]?)
\end{verbatim} 
\end{minipage} \\[-0.1em]\cmidrule(r){1-1}\addlinespace[-0.1em]
    \multirow{2}{*}{US} & 
\begin{minipage}{8.1in}
\begin{verbatim}
CCEVS-VR-((?P<cc>CC)-)?((?P<VID>VID)-?)?(?P<year>[0-9]{2})-(?P<counter>[0-9]+)
\end{verbatim} 
\end{minipage} \\
        &
\begin{minipage}{8.1in}
\begin{verbatim}
CCEVS-VR-((?P<cc>CC)-)?((?P<VID>VID)-?)?(?P<counter>[0-9]{4,5})(-(?P<year>[0-9]{4}))?
\end{verbatim} 
\end{minipage} \\\bottomrule
    \end{tabular}
    \vspace*{0.2cm}
    \caption{Regular expressions for Certificate IDs of Common Criteria schemes (with named groups in Python syntax).}
    \vspace*{-3cm}
    \label{tab:certid-regexes}
\end{table}
\end{landscape}

\end{document}